\documentclass[11pt,aps,article,showpacs,onecolumn,amsmath,amssymb,superscriptaddress]{revtex4-1}

\usepackage[letterpaper, margin=1.5cm]{geometry}
\usepackage{lmodern}
\usepackage{xcolor}
\usepackage[utf8]{inputenc}
\usepackage{xargs,lipsum,caption,changepage,ifthen}
\usepackage{amsmath}
\usepackage{amsfonts}
\usepackage{amssymb}
\usepackage{graphicx}
\usepackage{textcomp}

\usepackage{wrapfig}
\usepackage{floatrow}
\usepackage[utf8]{inputenc}
\usepackage[T1]{fontenc}
\usepackage{hyperref}

\usepackage{natbib}

\bibpunct{[}{]}{,}{s}{}{;}

\usepackage{mathtools}

\DeclarePairedDelimiter\abs{\lvert}{\rvert}%
\DeclarePairedDelimiter\norm{\lVert}{\rVert}%

\makeatletter
\let\oldabs\abs
\def\abs{\@ifstar{\oldabs}{\oldabs*}}
\let\oldnorm\norm
\def\norm{\@ifstar{\oldnorm}{\oldnorm*}}
\makeatother

\newcommand*{\dG}{\Delta G_{xx}}%

\renewcommand{\v}[1]{\textbf{#1}}
\newcommand{\tr}{\text{tr}}

\newcommand{\beginsupplement}{%
	\setcounter{table}{0}
	\renewcommand{\thetable}{S\arabic{table}}%
	\setcounter{figure}{0}
	\renewcommand{\thefigure}{S\arabic{figure}}%
	\setcounter{section}{0}
	\renewcommand{\thesection}{S\arabic{section}}%
	\setcounter{equation}{0}
	\renewcommand{\theequation}{S.\arabic{equation}}
	
}

\def\System{\textbf{Evolution of the topological and electronic properties of NI/TI heterostructures as the unit thickness of TI/NI layers is varied.} Top row: cross-section of the heterostructures, with t’ the thickness of TI unit layer and t that of NI, in QL. V’ and V are the corresponding hopping elements. Red and blue cones were used to represent the alternating helicity at adjacent interfaces. Middle row: electronic spectrum of the corresponding heterostructures, ranging from very thin NI limit, through the case of thick TI and NI layers, to very thin TI limit. Bottom row: weak antilocalization (WAL) channels arising and hybridizing as the relative unit layer thickness changes. The change in the number of self-intersecting loops can be compared to the corresponding behavior of Dirac cones across the setups in the top row. Red and blue loops are colored following the color code of interface states in the top row; the mixed color represents a coupled state.}

\def\TEMAFM{\textbf{TEM and AFM images of In$_{2}$Se$_{3}$/Bi$_{2}$Se$_{3}$ heterostructures.} 
\textbf{a}, TEM image of 4$\frac{8}{8}$ superlattice. \textbf{b}, Schematic of the same superlattice with labeled layers. \textbf{c}, AFM image, tapping mode, of the 4$\frac{8}{8}$ superlattice. Both the TEM and AFM images exhibit non-negligible roughness of overall morphology and interfaces. Nonetheless, the key transport properties, such as the number of WAL channels, are robustly determined by the number of unit cells and their thicknesses.}

\def\PhaseMain{\textbf{Theoretical phase diagrams in ($t$, $t'$)-plane (t: NI, t': TI)}. The diagrams are plotted for different sets of interface pairs, $i$ = 1 to $i$ = 10, based on a combination of the electronic spectrum (derived from Eq.~\eqref{eq:PhaseDiagcondition} in the supplement) and the theoretical WAL response of the superlattice (derived from Eq.~\eqref{eq:Combination}). \textbf{e} and \textbf{f} represent the effect of the Fermi level: low E$_F$ for \textbf{e} with the buffer layer and high E$_F$ for \textbf{f} without the buffer layer. The gray solid lines describe spectral transitions. In particular, at sufficiently large $i$, a trivial insulator (at $t' \rightarrow 0$, $t$ finite, white) is separated from a topological insulator (at $t \rightarrow 0$, $t'\gtrsim 5/i$) by a tongue-shaped cascade of transitions which anticipate the 3D Dirac semimetal occuring at $i \rightarrow \infty$. The color code represents the theoretical WAL coefficient $\tilde A_{\rm th}$. The labeled symbols in each phase diagram represent the samples being used in Fig. 3 and 4 for comparison with experiments. While the WAL calculation provides good agreement in most of the phase diagram space, it is only controlled to the top right of any gray line. A discussion of the applicability throughout the plot as well as details on the fitting of theoretical parameters are relegated to the supplement.}

\def\t{\textbf{Effective number of WAL channels, $\tilde A$ , for various NI/TI superlattices}. The $\tilde A$ values are obtained by fitting the magnetoconductance ($\Delta G_{xx}$) presented in the insets to the HLN formula. \textbf{a}, The 4$\frac{t}{8}$ series on low-E$_F$ substrates illustrates the strongly hybridized regime $t$ = 0 and $t$ = 1 where the system is electrically conducting as one 2D channel, a mixed regime at $t$ = 2, and a fully decoupled regime at $t \gtrsim$ 4. \textbf{b}, The \textit{i}$\frac{8}{t'}$ series on low-E$_F$ substrates with constant total Bi$_{2}$Se$_{3}$ thickness $i \cdot t’$ = 32 QL (red) and $i \cdot t’$ = 64 QL (black) demonstrate a nearly perfect one-to-one correspondence of $\tilde A$ to the number of interfaces pairs for $i \leq 4$. \textbf{c}, In the \textit{i}$\frac{t=t'}{t'}$ series on high-E$_F$ substrates with constant $i \cdot t'$, $\tilde A$ grows linearly while the unit layers are fully decoupled for $t’=t$ $\geq$ 8 QL, saturates for $t’=t = 6,5$ QL, and sharply drops toward 1 at $t’=t = 4$ QL.}

\def\Res{\textbf{Sheet resistance for metal-insulator transitions}. 
\textbf{a}, Resistance ($R_{xx}$) vs. temperature ($T$) plots. \textbf{b}, R$_{7K}$/R$_{min}$ extracted from \textbf{a}: when this value is larger than one, it implies the sample becomes insulating at low temperatures. All superlattices with at least 4 QL unit thickness are metallic. The $\frac{t=3}{t'=3}$ and $\frac{t=1}{t'=3}$ ($\frac{\text{NI}}{\text{TI}}$) films with high-E$_F$ remain metallic  at low $T$, whereas the same structures with low-E$_F$ become insulating. The $\frac{t=2}{t'=2}$ structures exhibit insulating behavior even for the high-E$_F$ case.}

\begin{document}
	
\title{Engineering artificial topological phases via superlattices}

\author{Pavel P. Shibayev}
\thanks{Equal contribution}
\author{Elio J. K{\"o}nig}%
\thanks{Equal contribution}
\affiliation{%
	Department of Physics and Astronomy, Rutgers, The State University of New Jersey, Piscataway, New Jersey 08854, U.S.A.\\
}%
\author{Maryam Salehi}%
\affiliation{%
	Department of Materials Science and Engineering, Rutgers, The State University of New Jersey, Piscataway, New Jersey 08854, U.S.A.\\
}%
\author{Jisoo Moon}%
\affiliation{%
	Department of Physics and Astronomy, Rutgers, The State University of New Jersey, Piscataway, New Jersey 08854, U.S.A.\\
}%
\author{Myung-Geun Han}%
\affiliation{%
	Center for Functional Nanomaterials, Brookhaven National Lab, Upton, New York 11973, U.S.A.\\
}%
\author{Seongshik Oh}%
\thanks{Correspondence should be addressed to: shibayev@alumni.princeton.edu, elio.j.koenig@gmail.com, ohsean@physics.rutgers.edu}
\affiliation{%
	Department of Physics and Astronomy, Rutgers, The State University of New Jersey, Piscataway, New Jersey 08854, U.S.A.\\
}%


\begin{abstract}
\textbf{The search for new topological materials and states of matter is presently at the forefront of quantum materials research. One powerful approach to novel topological phases beyond the thermodynamic space is to combine different topological/functional materials into a single materials platform in the form of superlattices. However, despite some previous efforts, there has been a significant gap between theories and experiments in this direction. Here, we provide the first detailed set of experimentally-verifiable phase diagrams of topological superlattices composed of archetypal topological insulator (TI), Bi$_{2}$Se$_{3}$, and normal insulator (NI), In$_{2}$Se$_{3}$, by combining molecular-level materials control, low-temperature magnetotransport measurements, and field theoretical calculations. We show how the electronic properties of topological superlattices evolve with unit-layer thicknesses and utilize the weak antilocalization effect as a tool to gain quantitative insights into the evolution of conducting channels within each set of heterostructures. This orchestrated study opens the door to the possibility of creating a variety of artificial-topological-phases by combining topological materials with various other functional building blocks such as superconductors and magnetic materials.}
\end{abstract}
\maketitle


In the past decade, there has been tremendous progress in understanding and synthesis of topological materials such as topological insulators, topological semimetals, and topological superconductors \citealp{m1,m2}. Many applications including spintronics, optoelectronics, and quantum computation have been conceptualized on these new materials platforms\citealp{m3,m4,m5}. So far, the majority of topological materials being investigated have been bulk crystals or thin films with well-defined topological characters\citealp{m1,m6,m7,m8}. A much less-explored approach is to combine different topological materials and create a new topological phase with distinct electronic properties that do not exist in each component, in the form of topological superlattices\citealp{m9,m10,m11,m12,m13}.

A major advantage of utilizing topological superlattices is the ease of tuning the materials parameters in LEGO-like fashion to design specific topological phases and functionalities. Theoretically, it was proposed that various topological phases including magnetic Weyl semimetals\citealp{m21} and Weyl superconductors\citealp{m22} that are hard to realize in single phase materials, could be implemented in topological superlattices by combining TIs with non-topological materials such as magnetic insulators or superconductors. Compared with the conceptual simplicity of topological superlattices, their experimental implementation is more challenging. Among other things, the non-topological layers should be structurally and chemically compatible with the van-der-Waals-bonded TI layers, which can be grown only at relatively low temperatures (200-300°C). There are not many materials with the desired functionalities that are still compatible with the TI layers. Among normal insulators, In$_{2}$Se$_{3}$ turns out to be well-compatible with the TI Bi$_{2}$Se$_{3}$ system\citealp{m12,m15,m13}, and they have been together used for TI/NI superlattices in a few previous studies\citealp{m16,m17,m19,m23}. Nonetheless, little is known as to whether and how their topological characters and electronic properties can be controlled.  Considering that TI/NI superlattices are physics-wise much simpler than the more exotic magnetic or superconducting topological superlattices, building a firm understanding of TI/NI superlattices is an essential step before digging into the complexity of other topological superlattices. Accordingly, the current study lays the foundation for designing artificial topological phases via superlattices, by providing the first realistic phase diagrams of TI/NI superlattices and comparing them with easily measureable transport properties. 

One of the unique approaches of the current study is that we take into account realistic Fermi levels when constructing the theoretical phase diagrams: this is critical because transport properties, such as weak antilocalization effects and metal insulator transitions, strongly depend on the Fermi levels. As for Bi$_{2}$Se$_{3}$ thin films, as previously shown by our group\citealp{namrata2012,S10,m24,moon2018}, the majority of the charge defects determining the Fermi level reside at the interface with the substrate. In Bi$_{2}$Se$_{3}$ films grown directly on Al$_{2}$O$_{3}$ substrates the Fermi level is above the conduction band minimum, whereas in films with In$_{2}$Se$_{3}$ buffer layers it is in the bulk band gap due to reduced interfacial defect density\citealp{m24}. Below we use both sets of samples, calling the first set by high-E$_{F}$ and the second by low-E$_{F}$. Although we focus more on the low-E$_{F}$ samples, we also study the high-E$_{F}$ samples to observe the effect of higher Fermi level. For all the heterostructures discussed below, we use the notation \textit{i}$\frac{t}{t'}$, where $i$ represents the number of super cells, $t’$ is the unit thickness of the TI Bi$_{2}$Se$_{3}$ layer, and $t$ is that of the NI In$_{2}$Se$_{3}$ layer, both in QL (quintuple layers). The details of film growth are provided in the Methods section.

We first start with a theoretical description of how the electronic/topological properties evolve with the thickness of each unit layer and construct a phase diagram. The results are summarized in Fig.~\ref{fig:System} and Fig.~\ref{fig:PhaseMain}. The experimental tuning knob is provided by the thicknesses of the two unit layers which determine the coupling of adjacent isolated topological surface states and enforce their progressive hybridization; see Fig.~\ref{fig:System}, top row. When the unit thickness is reduced from large values, Dirac interface states gradually lose their topologically protected metallic character, and eventually the bulk of the system becomes fully insulating. Depending on the relative thicknesses of TI and NI, overall topological or trivial insulating phases are realized at strong coupling. 

This qualitative picture is quantitatively reflected in a sequence of spectral transitions at which the number of Fermi surfaces consecutively decreases, shown in Fig.~\ref{fig:System}, central row, and which can be explained as follows. The hopping matrix elements $V$ and $V’$ generate a dispersion in stacking direction which in the present case of a finite system with open boundary conditions appears as \textit{i} twofold degenerate bands at positive energy (see Supplementary Section S1 for the exact analytical solution of a minimal model). As the hybridization gaps $\mathcal O(V,V’)$ of these bands cross the Fermi level $E_F$ one-by-one, Fermi surfaces sequentially disappear. These spectral transitions are plotted as solid lines in the phase diagrams of Fig.~\ref{fig:PhaseMain}. We highlight the leftmost of them, i.e. the metal-insulator transition (MIT), which (as all of the lines) strongly depends on the Fermi energy; see the comparison Fig.~\ref{fig:PhaseMain}(e) vs.~\ref{fig:PhaseMain}(f) for systems with low and high Fermi level, respectively. Furthermore, the appearance of a “tongue” of transitions at large $i$ and close to the low $t$ and $ t’$ represents the finite size precursor of the critical 3D Dirac semimetal separating topological and trivial insulators\citealp{m13}.

We perform low-temperature magnetotransport experiments in order to access these topological phase diagrams. The low-field magnetoresistance is dominated by the weak antilocalization (WAL) effect, the magnitude of which is generally universal and given by the number of \textit{distinct} 2D conduction channels. Physically, the origin\cite{Altshuler} of WAL is the constructive interference of amplitudes for clockwise and counterclockwise propagation of electrons along self-intersecting trajectories in the disordered conductor; see Fig.~\ref{fig:System}, bottom row. When all Dirac states of the superlattice are well-isolated, each interface contributes separately (central panel). However, when the rate of impurity-assisted decay to nearby surfaces is more frequent than the time needed to transverse the loop, interference effects of adjacent surfaces pair up (represented by mixed colors in Fig.~\ref{fig:System}) and the WAL continuously decreases. For a quantitative treatment of this effect we calculate the magnetoconductance $\sigma(B)$ perturbatively in weak tunneling $V, V’ \ll 1/\tau \ll E_F$ ($\tau$ is the elastic mean free time) employing an effective quantum field theory (non-linear sigma model). This leads to a generalization of the standard Hikami-Larkin-Nagaoka (HLN) formula\citealp{m28}, with a small field limit $\sigma(B) -\sigma(0) \stackrel{B \rightarrow 0}{\simeq} - \tilde A_{\rm WAL} (t,t') {B^2}/[{48 \pi B_\phi^2} ]$ ($B_\phi$ is the dephasing field). The full expression and the functional dependence of $\tilde A_{\rm WAL} (t, t’)$ are presented in Supplementary Section S1 and plotted in the color code of Fig.~\ref{fig:PhaseMain}. 
We conclude this theory section with a remark on the relevant energy scales of WAL crossover and spectral transitions. As mentioned above, the WAL crossover occurs when the decay rate to adjacent surfaces $V^2/(E_F^2 \tau)$ (or ${V’}^2/(E_F^2 \tau)$) reaches the temperature-dependent dephasing rate \cite{Altshuler} $1/\tau_\phi \sim T \ln(E_F \tau) /[E_F \tau]$, i.e. when $V, V’ \sim \sqrt{E_F T \ln (E_F \tau)}$, and thus at much weaker hopping than the spectral transitions, which occur at $V, V’ \sim E_F$ (see Fig.~\ref{fig:PhaseMain}). This is a reason why many thin TI slabs with spectroscopically clearly separated Dirac surface states act\citealp{mattSSC2015} as a single conduction channel in WAL experiments. 

In Fig.~\ref{fig:Figt}, we quantitatively compare the theoretical phase diagrams of Fig.~\ref{fig:PhaseMain} with transport experiments in terms of the effective number of WAL channels: low-E$_F$ samples for (a)-(b), and high-E$_F$ for (c). The main plot in Fig.~\ref{fig:Figt}(a) depicts the dependence of $\tilde A$ on $t$, with fixed $t’$ = 8 QL Bi$_{2}$Se$_{3}$ and $i$ = 4. These sets of films are also plotted in the corresponding $i$ = 4 phase diagram of Fig.~\ref{fig:PhaseMain}, along the $t'$ = 8 vertical trace. It shows that heterostructures of total Bi$_{2}$Se$_{3}$ thickness 32 QL with $t$ = 0 (no barriers) and $t$ = 1 (ultrathin 1 QL barriers) both have $\tilde A \approx$ 1, implying that the TI layers are strongly coupled across 1 QL-thick In$_{2}$Se$_{3}$. The case of $t$ = 2 yields $\tilde A \approx$ 3 (effectively three 2D channels), meaning that TI layers partially couple through the 2 QL In$_{2}$Se$_{3}$ layer. However, the cases of $t$ = 4 and $t$ = 8 both yield $\tilde A \approx$ 4, showing that the heterostructure splits into four isolated TI systems between $t$ = 2 and $t$ = 4 and remains so at higher values of $t$. The three regimes discussed above are easily distinguishable in the inset of Fig.~\ref{fig:Figt}(a), as the similar $\abs{\dG}$ values of the two structures in the connected regime ($t$ = 0, $t$ = 1) are well-separated from those of the mixed regime ($t$ = 2), separated further from the decoupled regime ($t$ = 4, $t$ = 8). 

In Fig.~\ref{fig:Figt}(b), we explore the ($t/t'$)-space of a generalized phase diagram along a constant NI unit thickness $t$ = 8 QL. Here, we vary the TI unit thickness (in multiples of 8 QL) for different sets of interface pairs while keeping fixed both the NI unit thickness (at $t$ = 8 QL) and the total Bi$_{2}$Se$_{3}$ thickness, $i \cdot t'$, at either 32 or 64 QL. These structures are plotted in the $i$ = 1, 2, 4, 8 phase diagrams of Fig.~\ref{fig:PhaseMain} along the $t$ = 8 horizontal traces. It clearly shows nearly one-to-one linear relationship between the $\tilde A$ value and $i$, the number of interface pairs, regardless of the total Bi$_{2}$Se$_{3}$ thickness. The one-to-one correspondence begins to deviate from the theory curve when $i$ reaches eight, which is probably because in such a large $i$ limit the system starts to evolve into a 3D regime, which cannot be well-handled by our effective 2D model.   

In Fig.~\ref{fig:Figt}(c), we keep each total thickness of the TI and NI layers constant at either 30 or 32 QL and measure their $\tilde A$ values while reducing their unit thickness ($t=t’$) and increasing the number ($i$) of interface pairs, ranging from 1$\frac{32}{32}$ to 10$\frac{3}{3}$. Up to 4$\frac{8}{8}$, the $\tilde A$ value grows almost linearly with $i$, slows down for 5$\frac{6}{6}$ and 6$\frac{5}{5}$, and then sharply drops for 8$\frac{4}{4}$ and 10$\frac{3}{3}$. It is notable that the TI unit thickness ($t’=6$ QL) where the WAL channels stop growing coincides with the point where the top and bottom surface states start to form a hybridization gap through strong coupling\citealp{m25}.  On the other hand, Figure 3(a) shows that 6 QL of NI unit is too thick to provide any measurable coupling for the WAL effect. Accordingly, the saturation of the number of effective WAL channels for 5$\frac{6}{6}$ and 6$\frac{5}{5}$ is mainly due to coupling through the TI layer rather than through the NI layer. For the large $i$ limit ($i=6,8,10$), our theory still aligns with the overall trend reasonably well despite not being controlled here; see Supplementary Section S1. 

Finally, in Fig.~\ref{fig:FigResist} we explore the metal-insulator transition (MIT), which corresponds to the leftmost spectral transition in each phase diagram of Fig.~\ref{fig:PhaseMain}. Among the samples we studied here, it is particularly interesting to compare two sets of samples in the context of MIT, 10$\frac{3}{3}$ and 10$\frac{1}{3}$ with low and high E$_F$. Those with high-E$_F$ remain metallic whereas those with low-E$_F$ become insulating at low temperatures, as can be seen in Fig.~\ref{fig:FigResist}(a) and (b). As for the high-E$_F$ case, the sample becomes insulating only if the unit thicknesses decrease to 2 as in 16$\frac{2}{2}$. The two sets of “i=10” samples belonging to the opposite sides of the MIT boundaries are well-captured in the phase diagrams, Fig.~\ref{fig:PhaseMain}(e) and (f).  However, considering that MIT is generally determined by the Ioffe-Regel criterion, $k_F l \approx 1$, where $k_F$ is the Fermi wave vector and $l$ is the mean free path, the exact condition for MIT should take into account not only the Fermi level but also the mean free path. Despite that, our simplified phase diagrams of Fig.~\ref{fig:PhaseMain} still capture the key aspects of transport properties. 

In conclusion, we have conducted a systematic study of topological superlattices consisting of alternating layers of Bi$_{2}$Se$_{3}$ and In$_{2}$Se$_{3}$ by combining field-theoretical calculations and low-temperature transport measurements. We have constructed a set of topological phase diagrams by calculating electronic spectrum and WAL parameters as a function of unit-layer thicknesses, also taking into account the Fermi levels. We have shown that these phase diagrams are well-matched with the transport measurements. Specifically, we have demonstrated that the WAL fitting parameter $\tilde A$ corresponds to the number of conducting channels and scales nearly linearly with the number of interface pairs. We have further demonstrated that as the unit thickness is reduced, the increased coupling between the surface states gradually weakens the topological protection, and the system eventually becomes fully insulating, while the exact MIT boundary is determined by sample details such as Fermi level and mean free path. Our study suggests that it is plausible to implement other more complex topological phases such as artificial 3D Dirac and Weyl semimetals from TI superlattices. The ability to design and manipulate topological properties using the unit layers of superlattices as building blocks can offer fascinating opportunities for new topological functionalities and device engineering. 
\clearpage

	{\includegraphics[scale=0.27]{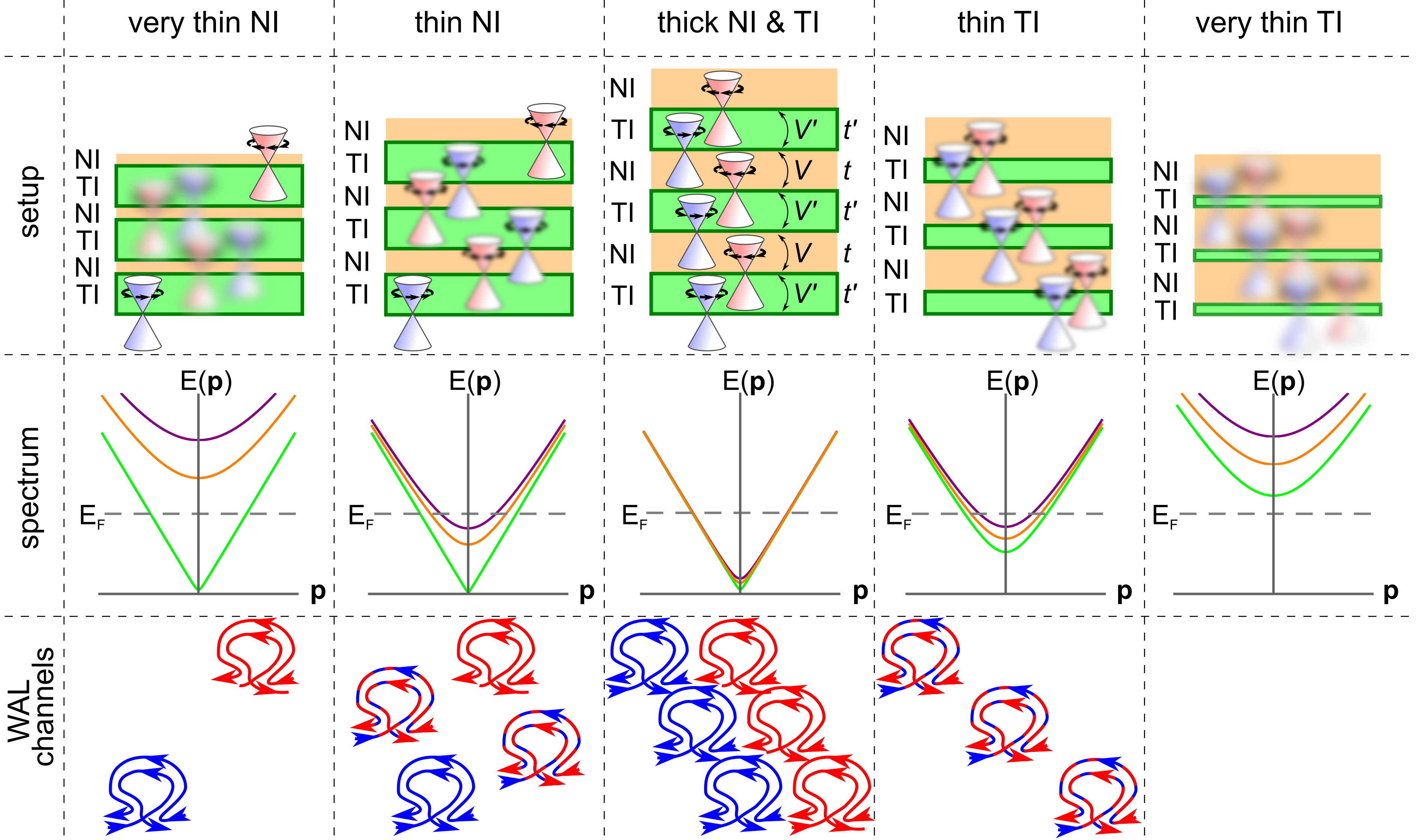}
	\captionof{figure}{\System}
	\label{fig:System}}

%


	{\includegraphics[scale=0.5]{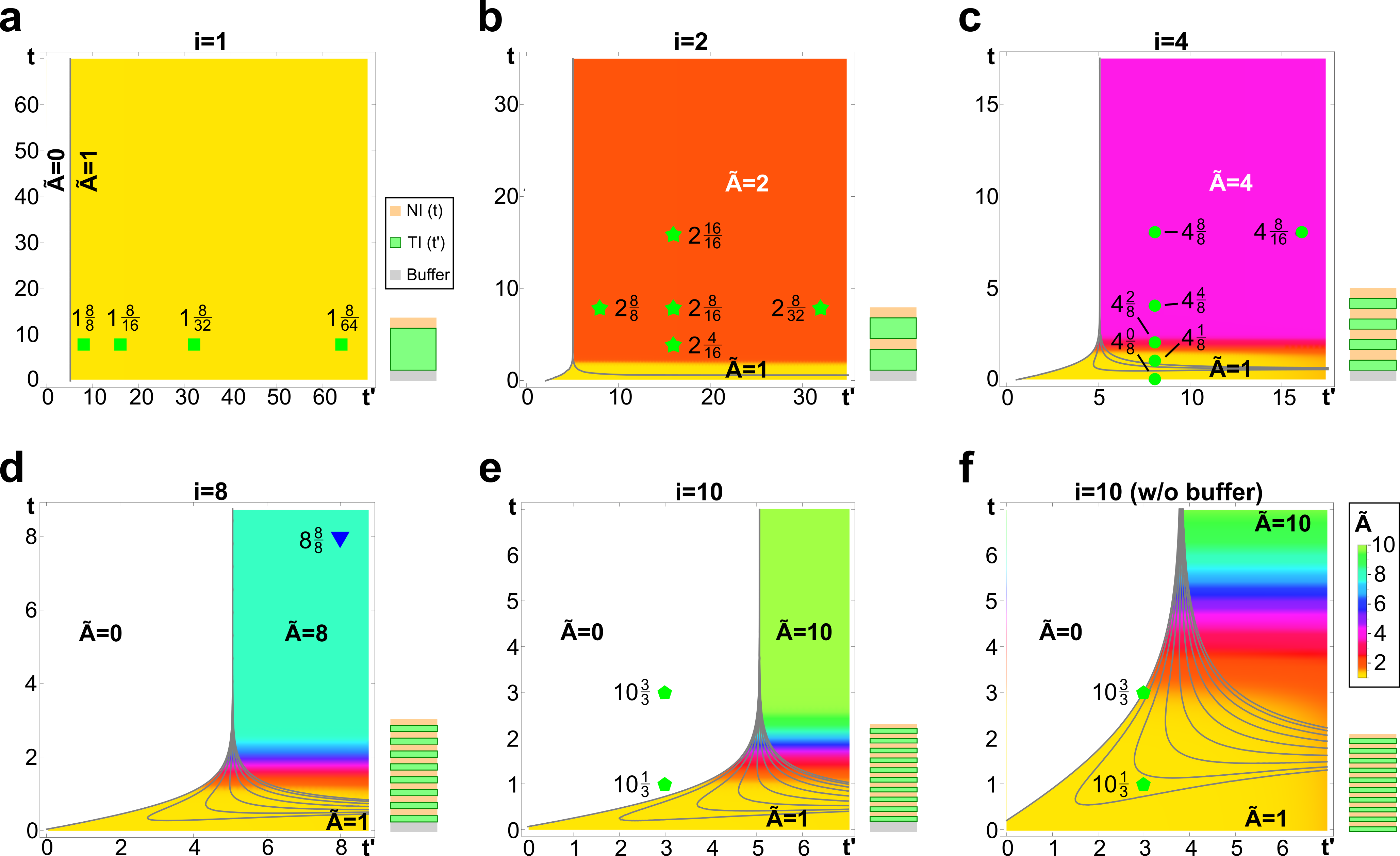}
	\captionof{figure}{\PhaseMain}
	\label{fig:PhaseMain}}

%


	{\includegraphics[scale=0.5]{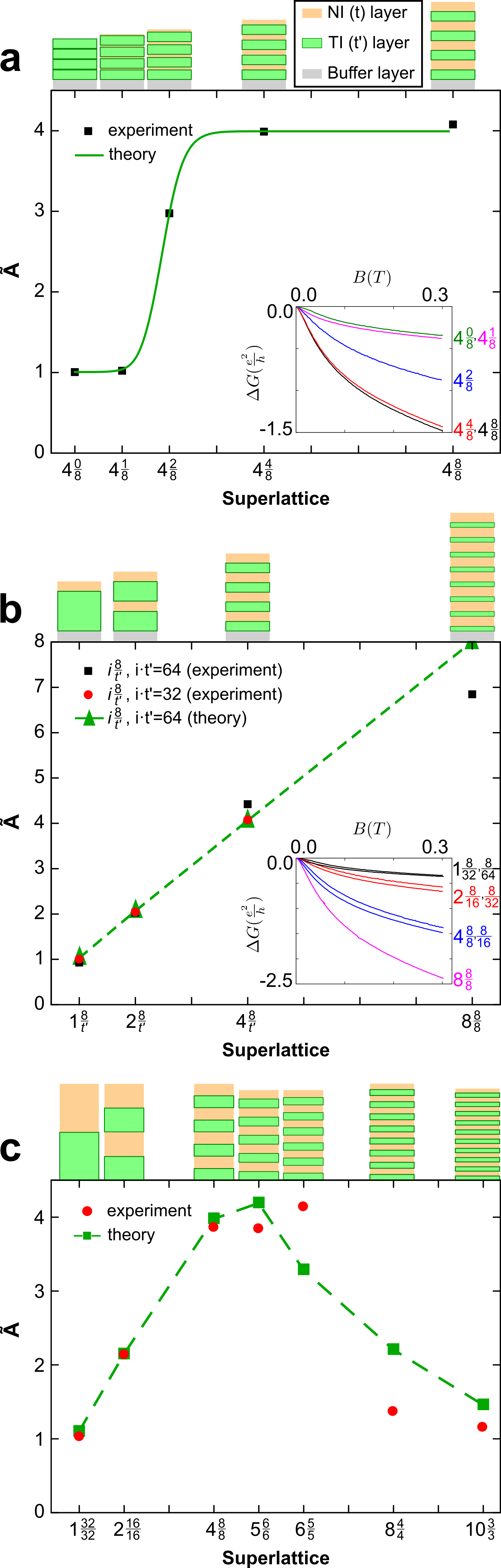}
	
	\captionof{figure}{\t}
	\label{fig:Figt}}

%


\clearpage
{\includegraphics[scale=0.5]{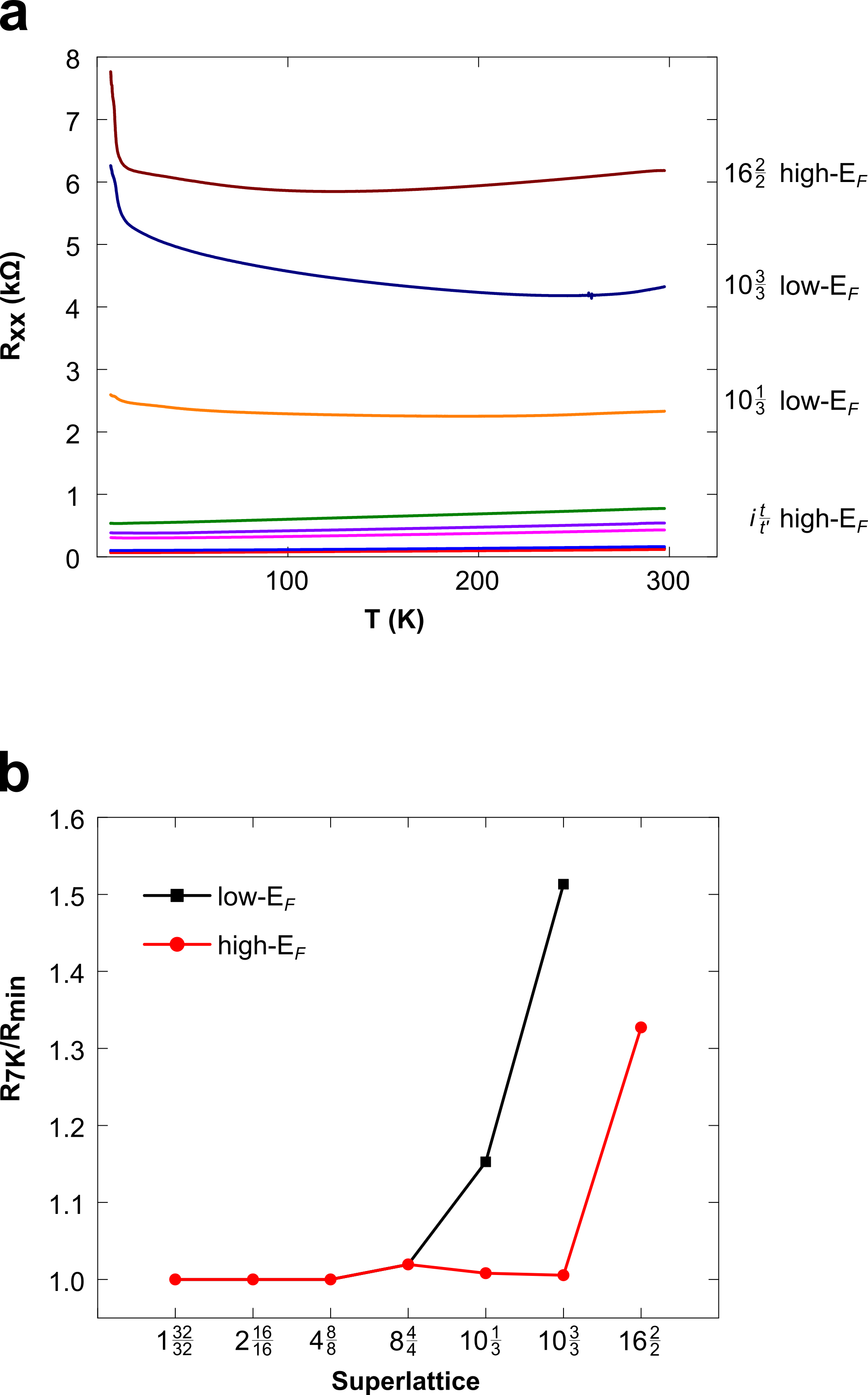}    
	\captionof{figure}{\Res}
	\label{fig:FigResist}}

%

\clearpage
\section*{Methods}
\subsection{Sample preparation}
The samples used in this study were prepared by a custom-designed molecular beam epitaxy (MBE) system (SVT Associates) on 10 × 10 mm$^2$ Al$_{2}$O$_{3}$(0001) substrates using a special buffer layer scheme, reported previously\citealp{m24} and briefly described as follows. We first deposit 3 QL of Bi$_{2}$Se$_{3}$ at 170°C, then heat the substrate to 300°C, deposit a buffer of 8 QL In$_{2}$Se$_{3}$, heat the system to 600°C, cool it to 270°C, and finally deposit the desired heterostructure via an automated procedure of alternating Bi$_{2}$Se$_{3}$ and In$_{2}$Se$_{3}$ layers at 270°C. All films were capped with an In$_{2}$Se$_{3}$ layer of thickness corresponding to the unit In$_{2}$Se$_{3}$ layer thickness within the heterostructure. The Bi, In, and Se fluxes were calibrated \textit{in situ} by a quartz crystal microbalance and \textit{ex situ} using Rutherford backscattering. Reflection high energy electron diffraction (RHEED) was used to monitor the epitaxial growth of the entire heterostructure. TEM and AFM images of a typical heterostructure are shown in Fig. S3 of the Supplementary Information.

\subsection{Transport measurements, background, calculations}
All transport measurements were performed with the standard Van der Pauw geometry in a closed-cycle cryostat of magnetic fields up to 0.6 T and temperatures down to 7.0 K. The films were briefly exposed to atmosphere during the transfer from the MBE to the measurement system; all samples were measured within a day of growth. Minor differences in substrate quality and transfer times accounted for the sample-to-sample variation in transport properties versus thickness shown in Figs.~\ref{fig:Figt} and~\ref{fig:FigResist}. In a typical resistance vs. magnetic field relationship the WAL effect is characterized by a cusp at small field values. This cusp can be most effectively observed by plotting the change of magnetoconductance (MC) as a function of magnetic field, $\Delta G_{xx}$ (B) ($B$ is varied up to 0.6 T in the present study).

In order to compute the WAL channel number, magnetoconductance data from transport measurements is fitted using the Hikami-Larkin-Nagaoka (HLN) formula, $\Delta G_{xx} (B) = \tilde A \frac{e^2}{2 \pi h} [\ln(\frac{B_\Phi}{B}) - \Psi (\frac{1}{2} + (\frac{B_\Phi}{B}))]$, where $h$ is Planck’s constant, $e$ is the electron charge, and $\Psi$ is the digamma function, by means of the fitting parameters $\tilde A$, which corresponds to the number of conducting 2D channels, and the dephasing field $B_{\Phi}$. It turns out that the channel number $\tilde A$ is very robust independent of sample qualities, whereas $B_{\Phi}$ can depend substantially on the film qualities. For example, single-slab undoped Bi$_{2}$Se$_{3}$ films beyond a critical thickness almost always exhibit $\tilde A$ value close to one\citealp{mattSSC2015}. As single-slab or heterostructure films pass through the metal-insulator transition (MIT), the WAL channel number $\tilde A$ collapses to zero.

\clearpage
\section*{Author contributions}
PPS grew the samples, carried out the transport and AFM measurements, and analyzed the data. EJK carried out the theoretical calculations and constructed the phase diagrams with inputs from PPS and SO. MS and JM assisted PPS for experiments.  MGH carried out the TEM measurement. SO supervised the project. PPS, EJK, and SO wrote the manuscript with inputs from the other coauthors. PPS, MS, JM, and SO are supported by Gordon and Betty Moore Foundation’s EPiQS Initiative (GBMF4418) and National Science Foundation (NSF) (EFMA-1542798). EJK is supported by DOE, Basic Energy Sciences grant DE-FG02-99ER45790. MGH is supported by the Materials Science and Engineering Divisions, Office of Basic Energy Sciences of the U.S. Department of Energy under Contract No. DESC0012704.

\clearpage
\beginsupplement
\section{Supplement: Theoretical determination of phase diagram}
\label{sec:Theory}

In this section of the supplementary materials we present the theoretical calculation of the phase diagrams underlying Fig.~\ref{fig:PhaseMain} of the main text. We base our calculation on a simple model Hamiltonian which contains topological surface states, intersurface hopping and disorder.

In Sec.~\ref{sec:ElectronicSpectrum}, the electronic spectrum of a clean system is determined. We find that, when the hopping elements are of the order of the Fermi energy $E_F$, a series of Lifshitz (i.e. Fermi surface topology changing) phase transitions occurs. 
In Sec.~\ref{sec:WAL}, the weak antilocalization (WAL) response of the superlattice is determined. As we show, Cooperon interference modes are sequentially gapped when the hopping elements are of the order $E_F l/l_\phi \ll E_F$ ($l$ is the elastic mean free path and $l_\phi$ the dephasing length). Therefore, in most but not all of the parameter space, the reduction of the WAL parameter precedes the Lifshitz transitions. 
We conclude that a combination of results from both sections is necessary to adequately describe the experiment. In particular, the (topological) transition from $\tilde A = 1$ to $\tilde A = 0$ at large hopping (thin superlattice layers) can only be explained as a spectral metal-insulator transition.

Planck's and Boltzmann's constants are both set to unity in this part of the supplementary materials.

\subsection{Electronic spectrum}
\label{sec:ElectronicSpectrum}

\subsubsection{Solution of superlattice tight binding problem}
As a first step, we determine the electronic spectrum in a clean system. To this end, we consider the following Hamiltonian for the superlattice structure which is schematically presented in Fig.~\ref{fig:System} of the main text 
\begin{equation}
	H = \int \frac{d^2p}{(2\pi)^2} \left [ \sum_{n=1}^{2i} (-)^n c^\dagger_n (\vec p) v \vec p \cdot \vec{\sigma} c_n(\vec p) + \sum_{n = 1}^{2i-1} V_n (c^\dagger_n(\vec p) c_{n + 1}(\vec p) + h.c)\right ].\label{eq:ModelHamiltonian}
\end{equation}
It should be considered as a minimal model that is reduced to the most important physics, i.e. Dirac surface states with kinetic Hamiltonian $\pm v \vec p \cdot \vec \sigma$ and scalar hopping elements between adjacent Dirac surface states $V_n = V'$ ($V_n = V$) for n odd (even). Here, $\vec \sigma$ are Pauli matrices and $\vec p$ is the momentum in the plane perpendicular to the stacking direction. 

In this section we determine the number $\tilde A_{\rm FS}$ of 2D Fermi surfaces of Eq.~\eqref{eq:ModelHamiltonian} as a function of $V'$ and $V$. 

For the diagonalization of Eq.~\eqref{eq:ModelHamiltonian} at arbitrary $i$ we switch to the notation $\bar V = \frac{V + V'}{2}$ and $\delta = \frac{V- V'}{V + V'}$ and to the eigenbasis of $\vec p \cdot \vec{\sigma}$. The eigenvalues $\lambda$ shall be measured in units of $\bar V$, than the eigenequations are
\begin{subequations}
	\begin{eqnarray}
		(1- \delta) x_2 &=& (\lambda +\epsilon_{\xi, \vec p}) x_1, \label{eq:EigenvalueEq1}\\
		(1-\delta) x_{2n-1} + (1+ \delta) x_{2n + 1} &=& (\lambda- \epsilon_{\xi, \vec p}) x_{2n},\label{eq:EigenvalueEq2} \\
		(1+\delta) x_{2n} + (1- \delta) x_{2n + 2} &=& (\lambda+\epsilon_{\xi, \vec p}) x_{2n+1},\label{eq:EigenvalueEq3} \\
		(1-\delta) x_{2i-1}&=& (\lambda-\epsilon_{\xi, \vec p}) x_{2i}. \label{eq:EigenvalueEq4}
	\end{eqnarray}
\end{subequations}
Here, $\epsilon_{\xi, \vec p} = \xi v \vert \vec p \vert/\bar V$ with $\xi = \pm 1$. Following loosely Refs. 1-2, 
the Ansatz for the non-normalized wave function $\psi = (x_1, \dots, x_{2i})$ is
\begin{equation}
	x_{2n -1} = \frac{1+\delta}{1-\delta} \sin((n-1)\theta) + \sin(n\theta); \quad x_{2n} = \frac{\lambda + \epsilon_{\xi, \vec p}}{1-\delta} \sin(n \theta). \label{eq:Ansatz}
\end{equation}
The eigenvalues $\lambda = \lambda_\pm(\vec p, \xi, \theta)$ are 
\begin{equation}
	\lambda_\pm(\vec p, \xi, \theta)   = \pm \sqrt{\epsilon_{\xi, \vec p}^2 + 2 [(1+\delta^2) + (1-\delta^2)\cos(\theta)]} \label{eq:Eigenenergies}
\end{equation} 
and the quantum number $\theta$ is to be determined. The Ansatz \eqref{eq:Ansatz} solves Eqs.~\eqref{eq:EigenvalueEq1},\eqref{eq:EigenvalueEq2},\eqref{eq:EigenvalueEq3} by construction. The concluding eigenequation \eqref{eq:EigenvalueEq4} yields the quantization condition
\begin{equation}
	(1- \delta)\sin((i+1)\theta)+(1+ \delta)\sin(i \theta) = 0. \label{eq:QuantizationCond}
\end{equation}
Analytical solutions to this equation for $i \leq 16$ were obtained using \textit{Mathematica} but are too cumbersome to be presented here. For $\delta < \delta_c = 1/(2i + 1)$ there are $i$ real angles $\theta \in (0, \pi)$ sufficing Eq.~\eqref{eq:QuantizationCond}, while for $\delta > \delta_c$, $i-1$ real solutions and one complex solution exist. The latter corresponds to the localized boundary mode, see Fig.~\ref{fig:TheoryCombinedPic} a) and b).

The value $\tilde A_{\rm FS}$ is determined by the number of states with energy $\lambda_\pm(\vec p, \xi, \theta) = E_F/\bar V$ for some $\vec p, \xi$. It follows that the phase boundaries are obtained by the family of conditions
\begin{equation}
	\lambda_\pm(0, +1, \theta) = E_F/\bar V \label{eq:PhaseDiagcondition}
\end{equation}
with $\theta$ being one of the $i$ solutions of Eq.~\eqref{eq:QuantizationCond}. In Fig.~\ref{fig:TheoryCombinedPic} c) we present the phase diagram for $i = 4$ in the $V' - V$ plane. 
\\
\\

\begin{center}
	{\includegraphics[scale=.5]{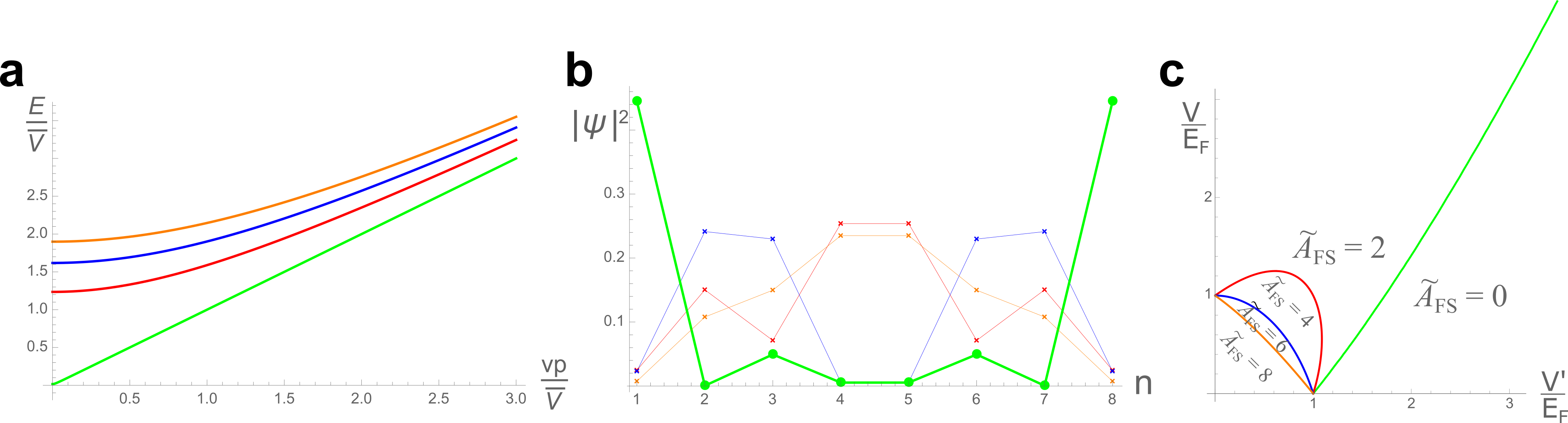} 
		\captionof{figure}{\textbf{Spectrum, wavefunctions and phase diagram of the minimal model Eq.~\eqref{eq:ModelHamiltonian} (here $i = 4$).} \textbf{a}, Positive spectrum $E = \bar V \lambda_{+1}(\vec p, \xi, \theta)$ and \textbf{b}, associated $\vec p = 0$ wavefunctions, both for $\delta = 0.5$.  Instead of the localized wave function (purple dots) a dispersive wave function exists for $\delta < 1/9$ (not shown). \textbf{c}, Phase diagram determined by the condition \eqref{eq:PhaseDiagcondition} in the ($V',V$) plane.}
		\label{fig:TheoryCombinedPic}}
\end{center}


\subsubsection{Asymptotic behavior, physical parameters}

Physically, the tight-binding approximation of well-defined surface states loses its meaning when the thicknesses $t$ or $t'$ approach zero. Nonetheless, a formal solution in this limit can be obtained and is presented here.
We anticipate that, assuming $V/ E_F \rightarrow \infty, V'/ E_F \rightarrow \infty$, on the transition between trivial and topological states, see Fig.~\ref{fig:TheoryCombinedPic} c), $V/V' \rightarrow \infty$ at any $i < \infty$. We prove this statement a posteriori.  
Expanding Eq.~\eqref{eq:PhaseDiagcondition} to leading order in $(1-\delta) \ll 1$ and employing the imaginary angle solution into Eq.~\eqref{eq:Eigenenergies} leads to
\begin{equation}
	\lambda {\simeq} \frac{V}{\bar V} \left (\frac{V'}{V}\right )^i.
\end{equation}
The transition line given by equating $\bar V \lambda = E_F$ is thus given by 
\begin{equation}
	V \simeq E_F \left (\frac{V'}{E_F}\right ) ^{\frac{i}{i - 1}}. \label{eq:Asymptote}
\end{equation}
We thus see that, in accordance with the assumption, $V/V' \rightarrow \infty$ on the transition line, the divergence is particularly rapid for $i$ not too large. Eq.~\eqref{eq:Asymptote} captures the numerical solution of Eqs.~\eqref{eq:Eigenenergies}, \eqref{eq:QuantizationCond} very well. 

We now turn to the functional behavior $V(t), V'(t')$. The following Ansatz is motivated by exponential localization of surface states in $z$ direction
\begin{subequations}
	\begin{eqnarray}
		V(t) &=& E_F e^{- \kappa (t - t_c)}, \\
		V'(t) &=& V_0' +E_F e^{- \kappa' (t' - t_c')}.
	\end{eqnarray}
	\label{eq:HoppingElements}
\end{subequations}
We introduced the thickness independent contribution  $V_0'$ to phenomenologically account for bulk mediated intersurface hybridization. At $V_0' = 0$, $t_c$ ($t_c'$) represent the critical thickness of transitions in the limit $t' \rightarrow \infty$ ($t \rightarrow \infty$). The mere existence of surface states implies $V_0' \ll E_F$. 
As mentioned, the $t \rightarrow 0$, $t' \rightarrow 0$ limits are at the border of applicability of the tight binding theory. To qualitatively account for this limit, large values of $V(0), V'(0)$ are needed, so that $\kappa t_c  \gtrsim 1, \kappa' t_c' \gtrsim 1$ is implied.

In the limit $t \rightarrow 0$ the NI layers are absent and a TI of total thickness $ t' i$ is realized. Thus the metal-insulator transition should occur at $t' = t'_c/i [1 + \mathcal O(V_0'/E_F)]$. Since $V(0) \gg E_F$ one can thus compare to Eq.~\eqref{eq:Asymptote}, which leads to
\begin{equation}
	\frac{V(0)}{E_F} {=}  \left (\frac{V'(t_c'/i)}{E_F}\right ) ^{\frac{i}{i - 1}} \Rightarrow \kappa t_c = \kappa' t_c ' + \mathcal O\left (\frac{V_0'}{E_F }e^{-\kappa' t_c' (1 - 1/i)}\right ). \label{eq:Condition}
\end{equation}
The fact that a solution to the left hand equation exists at all is a peculiarity of the particular functional form in Eq.~\eqref{eq:HoppingElements} and justifies the phenomenological choice. The implication on the right hand side restricts the parameters space to three independent fitting parameters. Expanding near $t = 0, t' = t_c'/i$ implies a linear relationship
\begin{equation}
	t \simeq \frac{i}{i-1} \frac{\kappa'}{\kappa} \left (t' - \frac{t_c'}{i}\right ). \label{eq:SuperLatticeAsymptote}
\end{equation}

\subsection{Weak antilocalization in a TI superlattice.}
\label{sec:WAL}

Weak (anti-)localization in superlattices and 3D anisotropic systems got certain attention over the years\citealp{WoelfleBhatt1984,SzottKirk1989,MauzWoelfle1997}, however open boundary systems\citealp{Novokshonov2007} were mostly disregarded and, to the best of our knowledge, TI specific calculations are not reported in the literature.
To derive the WAL response we consider the clean theory, Eq.~\eqref{eq:ModelHamiltonian}, supplemented by 
\begin{equation}
	H_{\rm dis} = \int d^2x \sum_{n = 1}^{2i} c_n^\dagger(\v x) U_n(\v x) c_n(\v x), \quad \langle U_n(\v x) U_{n'}(\v x) \rangle_{\rm dis} = \frac{1}{\pi \nu \tau} \delta(\v x - \v x') \delta_{nn'}. \label{eq:DisorderHamiltonian}
\end{equation}
We introduced the density of states $\nu =E_F/(2\pi v^2)$ at the Fermi level and the elastic scattering time $1/\tau$. For simplicity, the disorder potential is assumed to be Gaussian white noise distributed and without interlayer correlations or random interlayer hopping. As we discuss below, this minimal model captures all of the universal physics.

We follow the standard steps in the derivation of the low-energy diffusive theory (see e.g. Ref. 7
) and incorporate $V, V'$ perturbatively. Then, the non-linear sigma model action to second order in $V,V'$ takes the form (Einstein summation convention implied)
\begin{equation}
	S_\sigma = \frac{1}{16} \int d^2x \left \lbrace \sigma_{nn'} \tr [\nabla Q_n \nabla Q_{n'}] - m_{nn'} \tr Q_n Q_n'\right \rbrace. \label{eq:NLSMAction}
\end{equation}
Here, each $Q_n(\v x) = Q^T_n(\v x) = Q^{-1}_n (\v x)$ is a non-linear quantum field which takes values in the space of replicas, Matsubara frequencies and Nambu space and `$\tr$' is the trace in this space. Readers interested in details of the notation are referred to Ref. 8
, where an analogous notation was used. We derived the parameters $\sigma_{nn'}$ and $m_{nn'}$ microscopically to order $\mathcal O(V^2,V'^2, VV')$ using the standard semiclassical expansion in $1/E_F \tau \ll 1$. Concerning the prefactor of the gradient term, the knowledge that $\sigma_{nn'} = \mu \tau/2 \delta_{nn'} + \mathcal O((V \tau)^2,(V'\tau)^2)$ turns out to be sufficient for the WAL calculation. The mass term takes the form (an underscore denotes a matrix in the space of surface states $n = 1, \dots, 2i$)
\begin{equation}
	\underline m \simeq \frac{2\pi \nu}{\tau} \left (\begin{array}{ccccc}
		0 & \frac{V'^2}{E_F^2} & 0 & 0 & 0\\ 
		\frac{V'^2}{E_F^2} & 0 & \frac{V^2}{E_F^2} & 0 & 0 \\ 
		0 & \frac{V^2}{E_F^2} & 0 & \ddots & 0 \\ 
		0 & 0  & \ddots & 0 &\frac{V'^2}{E_F^2} \\
		0 & 0  & 0 &\frac{ V'^2}{E_F^2} &0 \\
	\end{array} \right ).
\end{equation}
The energy scale of the matrix elements can be readily understood: Due to opposite chirality, Fermi surface states of adjacent topological 2DEGs have vanishing overlap. Therefore, the leading order coupling stems from virtual processes involving the valence band and thus a factor $1/E_F^2$. This is specific to TIs and different in superlattices of conventional 2DEGs, see Ref. 4 
for comparison. 
Note that the form of Eq.~\eqref{eq:NLSMAction} is universal and only the precise numerical prefactors in $\underline \sigma$, $\underline m$ contain non-universal physics deriving from small length scales. In a perturbative calculation, interlayer disorder correlations and random interlayer hopping appear as mere corrections to $V^2$ and $V'^2$ at the sigma model level and can be readily incorporated.

Using linear response theory, the conductivity can be written as\citealp{KoenigMirlin2013,FootnoteSigmaCorrel}
\begin{equation}
	\sigma =  - \frac{\sigma_{nn'}}{8 \omega} 2\pi T\left \langle \tr[A_\omega, Q_n][A_{-\omega}, Q_{n'}] \right \rangle_{S_{\sigma}}, \text{ with } (A_\omega)_{\epsilon, \epsilon'}^{\alpha, \alpha'} = \left [\delta_{\epsilon -\epsilon', \omega} \frac{1 + \tau_y}{2} - \delta_{\epsilon -\epsilon', -\omega} \frac{1 - \tau_y}{2}\right ] \delta^{\alpha, \alpha'} \delta^{\alpha, 1}. \label{eq:KuboResponse}
\end{equation}
Here, the replica index is denoted $\alpha, \alpha'$, fermionic (bosonic) Matsubara frequencies $\epsilon, \epsilon'$ ($\omega$, it drops out in the low frequency limit), $T$ is the temperature and $\tau_y$ is the second Pauli matrix in Nambu space.

At the saddle point level, where $Q_n = \Lambda \; \forall n$ with $\Lambda_{\epsilon, \epsilon'} = \text{sign}(\epsilon)$, the conductivity is $\sigma_0 = \sum_{n} \sigma_{nn}$. Quadratic fluctuations about the saddle point have kinetics determined by Eq.~\eqref{eq:NLSMAction} and, at zero magnetic field, they are expressed in terms of the Cooperon kernel
\begin{equation}
	[\underline D^{-1}(\vec q)]_{nn'} = \frac{1}{4} (\sigma_{nn'} [\vec q^2+ l_\phi^{-2}] + \tilde m_{nn'}), \text{ where }  \tilde m_{nn'}:= \delta_{nn'} \sum_{\tilde n} m_{n\tilde n}-  m_{nn'},
\end{equation}
where we have introduced the dephasing length $l_\phi$ on phenomenological grounds. The quantum fluctuations lead to the WAL correction to conductivity
\begin{equation}
	\begin{split}
		\sigma - \sigma_0 = \frac{1}{2} \int \frac{d^2q}{(2\pi)^2} \text{sp} [\underline \sigma \underline D(\vec q)]= 2\int \frac{d^2q}{(2\pi)^2} \text{sp} [(\vec q^2+l_\phi^{-2} +l_{\tilde V}^{-2} \underline{M})^{-1}], \\
		\text{ with } \underline M = 2 \mathbf 1 -  \left (\begin{array}{ccccc}
			1+\Delta & 1-\Delta & 0 & 0 & 0\\ 
			1-\Delta & 0 & 1+\Delta & 0 & 0 \\ 
			0 & 1+\Delta & 0 & \ddots & 0 \\ 
			0 & 0  & \ddots & 0 &1-\Delta  \\
			0 & 0  & 0 &1-\Delta  &1+\Delta \\
		\end{array} \right ) .
	\end{split}
\end{equation}
We introduced $l_{\tilde V}^{-2}  = \frac{4}{l^2} \frac{V^2 + V'^2}{E_F^2}$ and $\Delta = \frac{V^2 - V'^2}{V^2 + V'^2}$ and the notation `$\text{sp}$' for the trace in the space of topological 2DEGs. It is convenient to switch to the eigenbasis of $\underline M$, we denote its eigenvalues as $\lambda_j(\Delta)$. The $\lambda_j(\Delta)$ are non-negative and a zero mode always exists, it reflects overall charge conservation. In a weak magnetic field, the $\vec q$ integral becomes a sum over Landau levels leading to ($B_\phi = 1/(4 e^2 l_\phi^2)$)
\begin{equation}
	\sigma(B) -\sigma( 0) = \sum_{j = 1}^{2i} Hi\left (\frac{B_\phi}{B}\left [1 + \frac{l^2_\phi}{l_{\tilde V}^2} \lambda_j(\Delta)\right ]\right ), \text{ with } Hi(x) = \frac{\ln(x) -\psi(x + 1/2)}{2\pi}.
\end{equation}
We define the theoretical value $\tilde A_{\rm WAL}$ by the low-field limit $\sigma(B) -\sigma( 0) \stackrel{B \rightarrow 0}{\simeq} - \tilde A_{\rm WAL} {B^2}/[{48 \pi B_\phi^2} ]$. Thus, theoretically,
\begin{equation}
	\tilde A_{\rm WAL} = \sum_{j = 1}^{2i} \left [1 + \frac{l^2_\phi}{l_{\tilde V}^2} \lambda_j(\Delta)\right ]^{-2}. \label{eq:AthWAL}
\end{equation}
This quantity is plotted in Fig.~\ref{fig:WALPhaseDiag}.
We remark that, for dephasing dominated by electron-electron interactions, 
${l^2_\phi}/{l_{\tilde V}^2} \sim ({V^2 + V'^2})/({E_F T \ln (E_F \tau)})$. Thus, $\tilde A_{\rm WAL}$ crosses over from $\tilde A_{\rm WAL} = 2i$ to smaller values already at scales $V, V' \sim \sqrt{{E_F T \ln (E_F \tau)}}$, i.e. at much smaller $V,V'$ as the Lifshitz transitions discussed in Sec.~\ref{sec:ElectronicSpectrum}. These values of $V, V'$ can be smaller than $1/\tau$, so that the perturbative derivation of Eq.~\eqref{eq:NLSMAction} is justified. 
\\
\\

\begin{center}
	{\includegraphics[scale=.35]{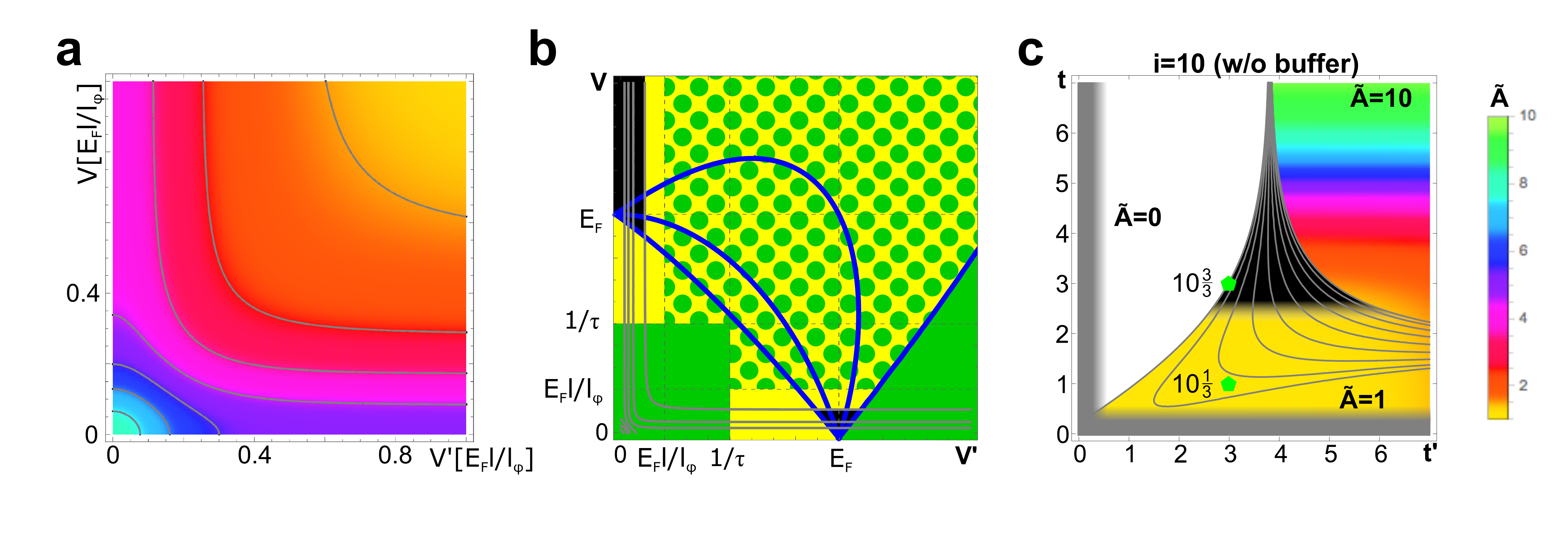} 
		\captionof{figure}{\textbf{Phase diagrams and applicability.} \textbf{a}, Phase diagram based on Eq.~\eqref{eq:AthWAL} in the (V, V') plane for the case $i = 4$. Topographic lines denote half integer values of $\tilde A_{\rm WAL}$. The asymmetry reflects the expectation $\tilde A_{\rm WAL} = 4$ ($\tilde A_{\rm WAL} = 5$) for $V' \gg V$ ($V \gg V'$) due to pairwise binding of Cooperon modes. \textbf{b}, Combination of the phase diagrams based on Eqs.~\eqref{eq:PhaseDiagcondition} (blue curves) and \eqref{eq:AthWAL} (black curves), again for $i = 4$. Panel b) also illustrates the regimes of validity of Eq.~\eqref{eq:Combination}. Controlled results are obtained in the green regions, in the green dotted region Eq.~\eqref{eq:Combination} is also asymptotically exact. The yellow regions can be expected to be qualitatively correct, while the Eq.~\eqref{eq:Combination} clearly fails in the black regions. \textbf{c}, Illustration of the regime of applicability of Fig. 2 of the main text in the exemplary case of samples with i = 10 and without buffer. The black wedge of inapplicability near $V' = E_F$ of panel b) translates to a black wedge near the metal insulator transition. At smallest t, t' the tight-binding approximation in stacking direction and the exponential dependence, Eq.~\eqref{eq:HoppingElements}, are unphysical (gray areas).}
		\label{fig:WALPhaseDiag}}
\end{center}

%
\subsection{Combination of weak antilocalization and spectral transitions.}
\label{sec:WALandES}

\subsubsection{Combined theory}

In order to describe the experimental observations and estimate the phase diagram for any value of $V, V'$, we use the following heuristic ansatz
\begin{equation}
	\tilde A_{\rm th} = \tilde A_{\rm WAL} \Theta(\tilde A_{\rm FS}). \label{eq:Combination}
\end{equation}
By definition we assume $\Theta(0) = 0$ for the Heaviside function. In Fig.~\ref{fig:WALPhaseDiag}, right, we justify the large regime of applicability of this expression. In particular, Eq.~\eqref{eq:Combination} is controlled in the regime of a trivial band insulator and in the regime $V, V' \ll 1/\tau$. In the regime $V, V' \gg E_F l/l_\phi$ the WAL calculation predicts $\tilde A_{\rm WAL} \simeq 1$, this is the minimum value ensured by overall particule number conservation. Assuming monotonicity, it immediately follows that Eq.~\eqref{eq:Combination} provides the asymptotically correct answer in the regime $V, V' \gg E_F l/l_\phi$ (represented by green dots in Fig.~\ref{fig:WALPhaseDiag}). The yellow regions, in particular in the phase $\tilde A_{\rm FS} = 2i$, can be expected to be qualitatively correct even for $V>1/\tau$ or $V'>1/\tau$, however results are not controlled. In the red regions, Eq.~\eqref{eq:Combination} the heuristic ansatz \eqref{eq:Combination} clearly provides wrong answers. 

\subsubsection{Fit to experiment}

In order to fit the five parameters $V_0'/E_F, \kappa', \kappa, t_c', t_c$ entering Eq.~\eqref{eq:HoppingElements} and thus Eq.~\eqref{eq:Combination} to the experimental data for all samples with a buffer layer we minimize the root mean distance
\begin{equation}
	\text{RMSD} = \sqrt{\sum_\alpha (\tilde A_{\rm th} - \tilde A_{\rm exp})^2/\sum_\alpha 1}.
\end{equation}
We impose the condition $\kappa t_c = \kappa't_c'$, see Eq.~\eqref{eq:Condition}, and use $l_\phi/l = 4$ in accordance with our fits of the HLN magnetoresistence. A first coarse search of the five dimensional parameter space helps to discriminate the regime $(V_0'/E_F, \kappa', t_c', t_c) \in [0.1,0.2]\times [0.2,0.3] \times  [4.5, 4.6] \times [0.55, 0.65]$ to be most relevant (all lengths in QL). Within this subspace we then perform a variation in steps of $0.01$ which reveals a minimum at $(V_0'/E_F, \kappa', t_c', t_c) = (0.13, 0.29, 4.58, 0.6)$ with a positive definite variation matrix
\begin{equation}
	\text{RMSD} \simeq 0.604 +\left (
	\delta V_0'/E_F ,
	\delta \kappa' ,
	\delta t_c ,
	\delta t_c'
	\right ) \left(
	\begin{array}{cccc}
		0.82 & -0.12 & 0.002 & 0.02 \\
		-0.12 & 16.39 & 1.02 & -11.10 \\
		0.002 & 1.02 & 0.07 & -0.71 \\
		0.02 & -11.10 & -0.71 & 7.79 \\
	\end{array}
	\right)\left (\begin{array}{c}
		\delta V_0'/E_F \\ 
		\delta \kappa' \\ 
		\delta t_c \\ 
		\delta t_c'
	\end{array} \right ).
\end{equation}

For the fit to the data set without buffer we exploit that the major change as compared to the previous data set is the level of the Fermi energy affecting of $V_0', \kappa, \kappa'$, while exponential prefactors $e^{\kappa' t_c'}$ and $e^{\kappa t_c}$ are expected to be $E_F$ independent. We thus use the result for systems with buffer as an Ansatz for hopping in units of $E_F$ 
\begin{equation}
	V' = 0.13/ \bar E_F + e^{-0.29 (\kappa' t' - 
		4.58)}, V=  e^{-(0.29) 4.58/0.6 (\kappa t -0 .6)}.
\end{equation}
Again, a coarse search is followed by a fine search in  steps of $0.01$, now in the regime $(\bar E_F, \kappa', \kappa) \in [1.2, 1.4] \times [1.25, 1.35]\times [ 0.35, 0.45]$. The minimum at $(1.39, 1.29, 0.36)$ is characterized by 

\begin{equation}
	\text{RMSD} \simeq 0.60 + \left (\delta E_F, \delta \kappa', \delta \kappa \right )\left(
	\begin{array}{ccc}
		0.016 & 0.017 & 0.013 \\
		0.017 & 1200.77 & 0.10 \\
		0.013 & 0.10 & 1409.94 \\
	\end{array}
	\right)\left (\begin{array}{c}
		\delta E_F \\ 
		\delta \kappa' \\ 
		\delta \kappa
	\end{array} \right ).
\end{equation}

%
%

\clearpage
\section{Supplement: TEM and AFM Characterization}

TEM and AFM images of our MBE-grown typical In$_{2}$Se$_{3}$/Bi$_{2}$Se$_{3}$ superlattices are shown for the 4$\frac{8}{8}$ case in Fig.~\ref{fig:TEMAFM}. TEM samples are prepared by focused-ion-beam (FIB) with 5 keV Ga$^{+}$ ions to minimize ion-beam-induced damages. For high-angle annular dark-field (HAADF) scanning transmission electron microscopy (STEM), a JEOL ARM 200CF equipped with a cold field-emission gun and double-spherical aberration correctors at the Brookhaven National Laboratory operated at 200 kV was used with the range of collection angles from 68 to 280 mrad. AFM acquisition was done using a Bruker Multimode 8 AFM in tapping mode with a cantilever of stiffness in the range 10-70 N/m at the Department of Chemistry in Fordham University.

\begin{center}
	{\includegraphics[scale=1.4]{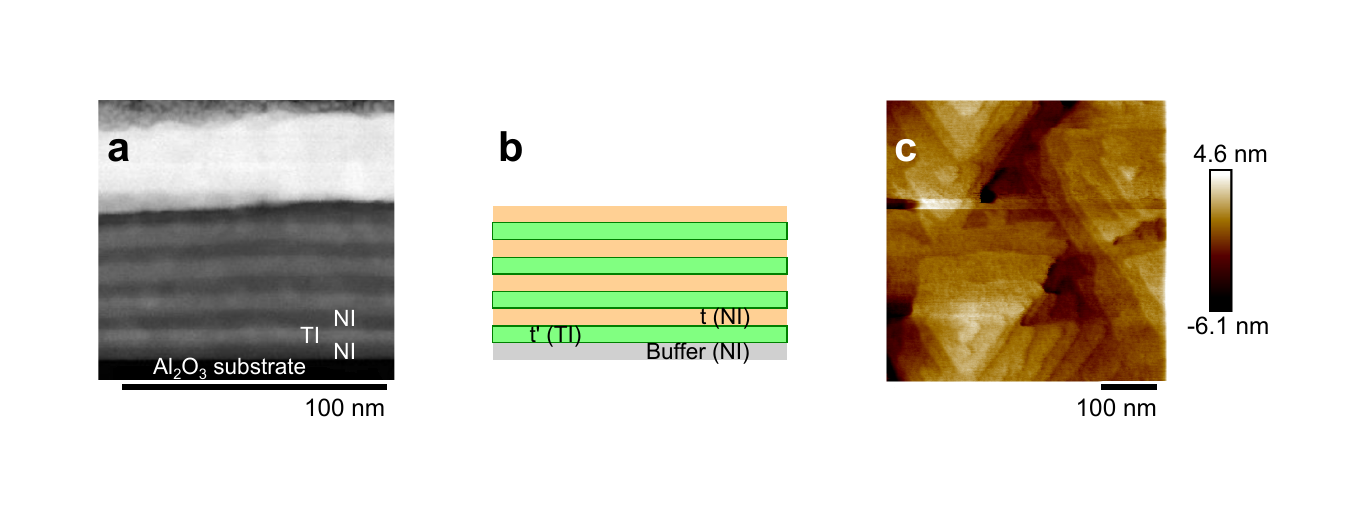}
		\captionof{figure}{\TEMAFM}
		\label{fig:TEMAFM}}
\end{center}

\clearpage
\section*{Supplement References}


\begin{thebibliography}{9}
\bibitem{FootnoteSigmaCorrel} Formally, the response to the $U(1)$ gauge potential also involves a non-local correlator denoted as $B_2$ in Ref. 8. Its contributions to semiclassical and WAL response vanish in the replica limit.
\end{thebibliography}
\end{document}